\documentclass{mem}
\usepackage{natbib}\usepackage{txfonts}\usepackage{balance}
\usepackage{flushend}
\usepackage{graphicx}
\usepackage[breaklinks,pdftex]{hyperref}
\idline{75}{282}
\begin{document}
\def\teff{$T\rm_{eff }$}
\def\kms{$\mathrm {km s}^{-1}$}

\title{
GRB multi-TeV detection: \\ Beyond standard physics?
}


\author{
G.\,Galanti\inst{1} 
\and L.\,Nava\inst{2,3}
\and M.\,Roncadelli\inst{4,2}
\and F.\,Tavecchio\inst{2}
\and G.\,Bonnoli\inst{2}
          }

\institute{
INAF, Istituto di Astrofisica Spaziale e Fisica Cosmica di Milano, Via Alfonso Corti 12, I -- 20133 Milano, Italy,
\email{gam.galanti@gmail.com}
\and
INAF, Osservatorio Astronomico di Brera, Via Emilio Bianchi 46, I -- 23807 Merate, Italy
\and
INFN, Sezione di Trieste, Via Alfonso Valerio 2, I -- 34127 Trieste, Italy
\and
INFN, Sezione di Pavia, Via Agostino Bassi 6, I -- 27100 Pavia, Italy
}

\authorrunning{Galanti et al.}

\titlerunning{GRB multi-TeV detection: Beyond standard physics?}

\date{Received: Day Month Year; Accepted: Day Month Year}

\abstract{
The recent detection by LHAASO up to 18 TeV of the gamma ray burst GRB 221009A at redshift $z = 0.151$ challenges standard physics because of the strong absorption due to the extragalactic background light (EBL) for photons with energies above 10 TeV. Emission models partially avoiding EBL absorption proposed to explain such an event are unsatisfactory since they require peculiar and contrived assumptions. By introducing in magnetized media the interaction of photons with axion-like particles (ALPs) -- which are a generic prediction of most theories extending the standard model of particle physics towards a more satisfying theory -- the detection of GRB 221009A can be naturally explained, thereby providing a strong hint at ALP existence. 
}
\maketitle{}

\section{Introduction}

The detection of the Gamma Ray Burst GRB 221009A at very-high-energy (VHE, ${\cal E} > 100 \, {\rm GeV}$) by the~\citet{LHAASO} with the highest photon energy reaching $18 \, \rm TeV$ and less reliably by the~\citet{carpet} up to $251 \, {\rm TeV}$ challenges standard physics expectations: at the redshift $z = 0.151$ of  GRB 221009A measured by the~\citet{redshift}, photons with energy ${\cal E} >10 \, \rm TeV$ are extremely difficult to detect because of their interaction with the extragalactic background light (EBL) photons~(\citet{dwek}).

In particular, the attenuation of photons with ${\cal E} = 18 \, \rm TeV$ is at least ${\cal O} (10^6-10^8)$, as shown by the several existing EBL models (see e.g.,~\citet{dominguez,gilmore,franceschinirodighiero,saldanalopez}). The luminosity required to justify the observed VHE photons from GRB 221009A turns out to be 
excessively high within all conventional GRB models.

In order to find a solution to such a problem, in~\citet{ALPinGRB} it is proposed that photons oscillate into axion-like particles (ALPs) in the presence of external magnetic fields permeating the media crossed by the beam (see also~\citet{ALPinGRB2}). As a result, the effective photon optical depth gets reduced, since when the beam photons behave as ALPs they are {\it not} EBL absorbed. This scenario has been deeply investigated by us in the last years (\citet{universe}).

In~\citet{ALPinGRB} we also analyze the effects on GRB 221009A produced by the Lorentz invariance violation (LIV).

\section{Standard view}

VHE photons originated at cosmological distances suffer absorption caused by their interaction with the EBL, namely photons emitted by stars and possibly reprocessed by dust during the whole Universe history.

All existing EBL models (see e.g.,~\citet{dominguez,gilmore,franceschinirodighiero,saldanalopez}) are quite consistent among each other in the optical-ultraviolet range, but they differ in the infrared band which is responsible for photon absorption above $\sim 5 \, \rm TeV$.

As based on observations from space, thereby avoiding the zodiacal light problem, the most robust EBL model appears to be that by~\citet{saldanalopez}. Therefore, we hereafter employ the~\citet{saldanalopez} EBL model which predicts -- within conventional physics -- a photon survival probability amounting to $P_{\rm CP}\simeq 1 \times 10^{-8}$ at ${\cal E} = 18 \, \rm TeV$ and $P_{\rm CP}\simeq 3 \times 10^{-6}$ at ${\cal E} = 15 \, \rm TeV$. The latter energy value is chosen in order to take into account an uncertainty of $(15-20)\%$ concerning the detection by the~\citet{LHAASO}. Regarding the possible detection by the~\citet{carpet}, the model by~\citet{saldanalopez} predicts $P_{\rm CP}\sim 0$ both at $251 \, {\rm TeV}$ and at $100 \, {\rm TeV}$, with the latter energy value chosen to account for an uncertainty of $\sim 50 \%$.

Because of the very low values assumed by $P_{\rm CP}$, photons with ${\cal E} > 10 \, \rm TeV$ are strongly absorbed and their detection requires a very high TeV luminosity, which is very difficult to explain within conventional emission models.

\section{Axion-like particles}

In order to solve the tension between the observation at VHE of GRB 221009A and conventional emission models, we invoke the possibility that photons oscillate into axion-like particles (ALPs).

ALPs are predicted by superstring and superbrane theories (see e.g.,~\citet{JR2010,r2012}) and turn out to be extremely light neutral pseudo-scalar bosons with mass $m_a$  interacting with photons with coupling $g_{a \gamma \gamma}$ through the Lagrangian 
\begin{equation}
{\cal L}_{a \gamma \gamma} = - \, \frac{1}{4} \, g_{a \gamma \gamma} \, F_{\mu \nu} \, {\tilde{F}}^{\mu \nu} \, a = g_{a \gamma \gamma} \, {\bf E} \cdot {\bf B} \, a~,
\label{a2}
\end{equation}
where $a$ represents the ALP field, while $F_{\mu \nu}$ (with dual ${\tilde{F}}^{\mu \nu}$) is the electromagnetic tensor whose electric and magnetic components are ${\bf E}$ and ${\bf B}$, respectively. The most solid bound on the ALP parameter space $(m_a, g_{a\gamma\gamma})$ is represented by $g_{a \gamma \gamma} < 0.66 \times 10^{- 10} \, {\rm GeV}^{- 1}$ for $m_a < 0.02 \, {\rm eV}$ at $2 \sigma$ level by the~\citet{cast}, while the most stringent ones read: (i) $g_{a \gamma \gamma} < 6.3 \times 10^{- 13} \, {\rm GeV}^{- 1}$ for $m_a < 10^{- 12} \, {\rm eV}$ by~\citet{limJulia} and (ii) $g_{a \gamma \gamma} < 5.4 \times 10^{- 12} \, {\rm GeV}^{- 1}$ for $m_a < 3 \times 10^{- 7} \, {\rm eV}$ by~\citet{mwd}. The QED vacuum polarization (\citet{hew1,hew2,hew3}), and the photon dispersion on the CMB (\citet{raffelt2015}) effects must also be considered in this context.

In the presence of an external magnetic field ${\bf B}$, photons represented by ${\bf E}$ in Eq.~(\ref{a2}) can convert into ALPs producing: (i) photon-ALP oscillations~(\citet{mpz,raffeltstodolsky}) and (ii) change of the polarization state of photons~(\citet{mpz,raffeltstodolsky}). As a result, photon-ALP interaction produces many consequences both on observed astrophysical spectra (see e.g.,~\citet{drm2007,dgr2011,simet2008,sanchezconde2009,trgb2012,trg2015,wb2012,kohri2017,gtre2019,grdb}) and on the detected photon polarization (see e.g.,~\citet{ALPpol1,bassan,ALPpol2,ALPpol3,ALPpol5,day,galantiTheo,galantiPol,grtcClu,grtBlazar}).

In GRB 221009A VHE photons are produced in the GRB jet and they can then oscillate into ALPs inside all crossed magnetized media (the GRB jet, the galaxy hosting the GRB, the extragalactic space, and the Milky Way). As demonstrated in~\citet{ALPinGRB}, the photon-ALP conversion inside the GRB jet is negligible since the propagation length of the photon-ALP beam inside the source is too small to produce an efficient photon-ALP conversion. Following the results presented in~\citet{GRB221009Ahost}, the galaxy hosting GRB 221009A is a disc-like one observed edge-on with the GRB located close to the nuclear region. Therefore, we use previous information to compute the photon-ALP conversion in the host, and we consider the two most likely possibilities concerning the nature of the host galaxy: (i) a typical spiral (see e.g.~\citet{SpiralBrev}) and (ii) a starburst with intermediate features similar to M82~(\citet{Thompson2006,LopezRodriguez2021}). Photon-ALP conversion inside the host is efficient in the case of both a spiral and a starburst galaxy (see~\citet{ALPinGRB}). Concerning photon-ALP interaction inside the extragalactic space, we follow the results reported in~\citet{galantironcadelli20118prd,grjhea} by considering both an efficient photon-ALP conversion with an extragalactic magnetic field strength $B_{\rm ext} = 1 \, \rm nG$ (more likely, see e.g.,~\citet{reessetti1968,hoyle1969,kronbergleschhopp1999,furlanettoloeb2001}) and a negligible photon-ALP conversion with $B_{\rm ext} < 10^{-15} \, \rm G$ (extremely conservative, see e.g.,~\citet{neronovvovk,durrerneronov,upbbext}). Finally, we use the magnetic field map by~\citet{jansonfarrar1,jansonfarrar2} and the electron number density model by~\citet{yaomanchesterwang 2017} in order to evaluate the photon-ALP conversion in the Milky Way. Once the transfer matrices of the photon-ALP beam in all the previous regions are evaluated and combined, we obtain the photon survival probability in the presence of photon-ALP oscillations $P_{\rm ALP}$, as discussed in~\citet{ALPinGRB}.

\begin{figure}
\resizebox{\hsize}{!}{\includegraphics[clip=true]{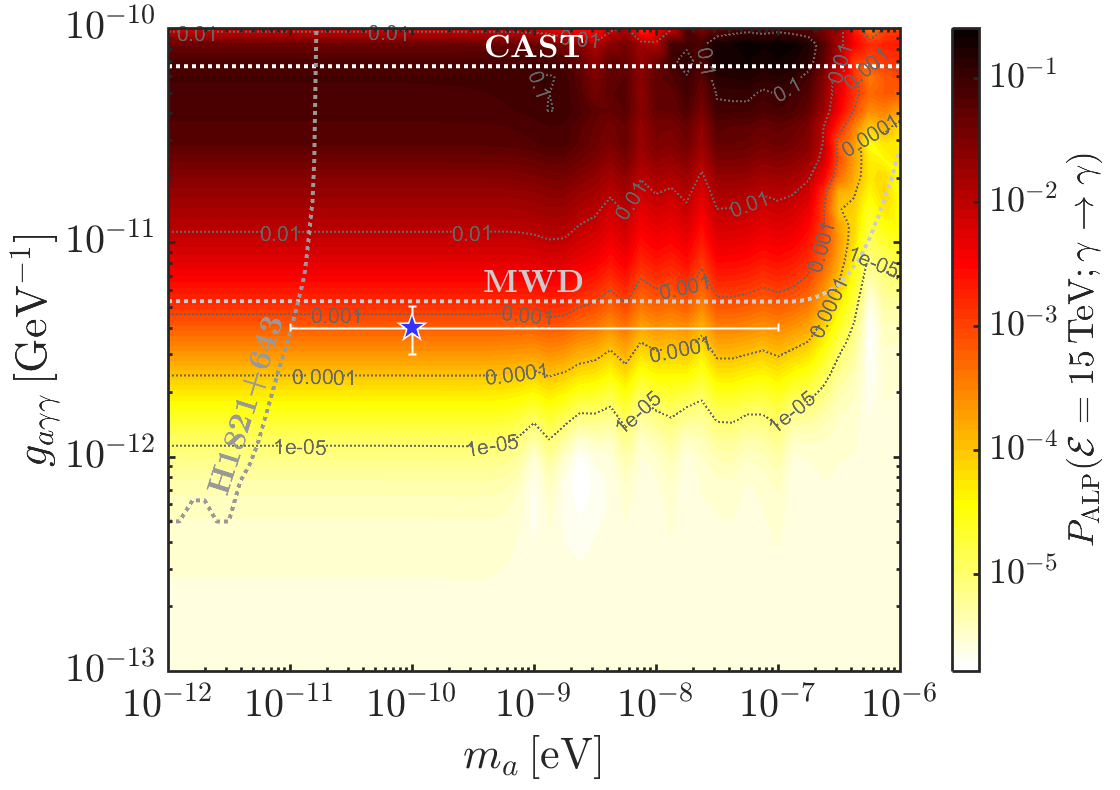}}
\caption{
\footnotesize
Behavior of $P_{\rm ALP}$ at ${\cal E} = 15 \, \rm TeV$ as a function of $m_a$ and $g_{a\gamma\gamma}$ by assuming the EBL model of~\citet{saldanalopez}, $B_{\rm ext}=1 \, \rm nG$ and a starburst hosting galaxy. ALP bounds are also plotted (\citet{cast,limJulia,mwd}). (Credit:~\citet{ALPinGRB}).}
\label{parSpaceStarburst}
\end{figure}

In Fig.~\ref{parSpaceStarburst} we plot $P_{\rm ALP}$ at ${\cal E} = 15 \, \rm TeV$ as a function of $m_a$ in the range $10^{- 12} \, {\rm eV} \lesssim m_a \lesssim 10^{- 6} \, {\rm eV}$ and of $g_{a\gamma\gamma}$ in the interval $10^{- 13} \, {\rm GeV}^{- 1} \lesssim g_{a \gamma \gamma} \lesssim 10^{- 10} \, {\rm GeV}^{- 1}$ by assuming the EBL model of~\citet{saldanalopez}, $B_{\rm ext}=1 \, \rm nG$ and a starburst hosting galaxy.

\begin{figure}
\resizebox{\hsize}{!}{\includegraphics[clip=true]{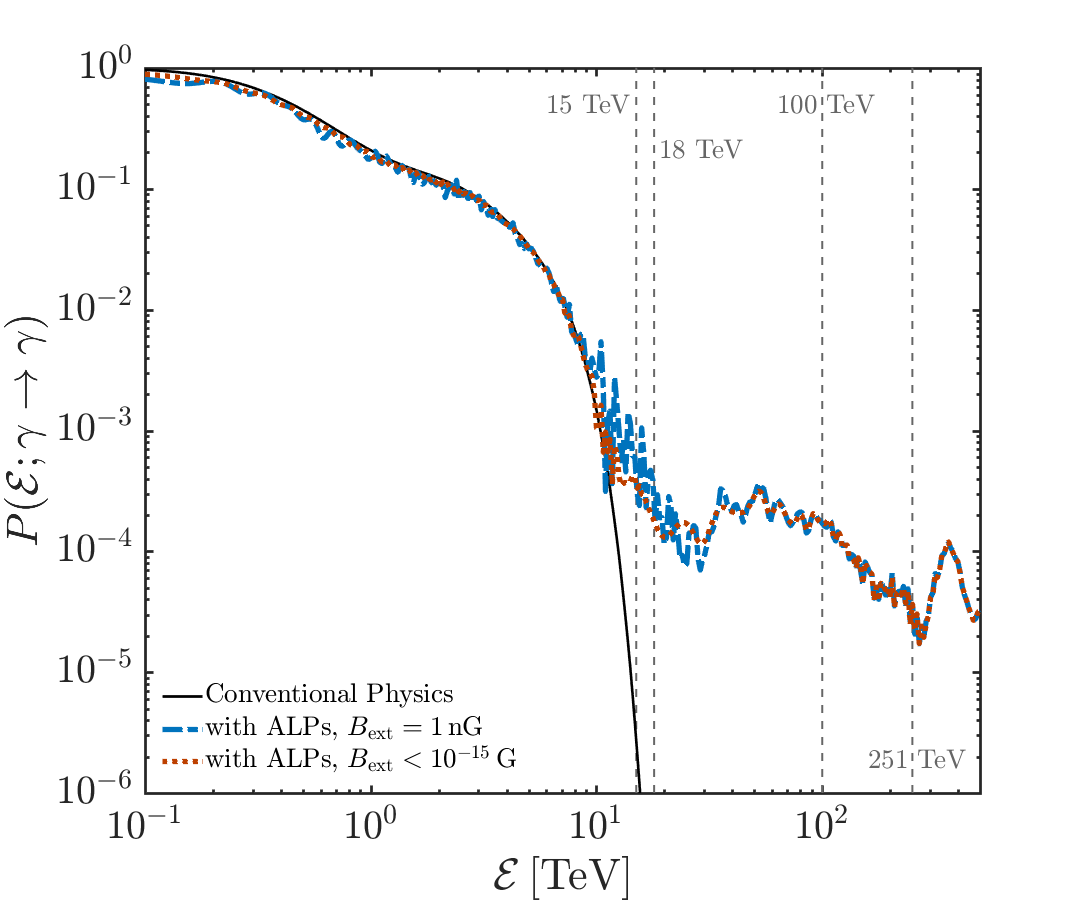}}
\caption{
\footnotesize
Behavior of $P_{\rm CP}$ and of $P_{\rm ALP}$ (for $B_{\rm ext} = 1 \, \rm nG$ and $B_{\rm ext} < 10^{-15} \, \rm G$) versus ${\cal E}$ for the case of a starburst hosting galaxy and employing the EBL model of~\citet{saldanalopez}. We assume $m_a=10^{-10} \, \rm eV$ and $g_{a\gamma\gamma}=4 \times 10^{-12} \, \rm GeV^{-1}$. (Credit:~\citet{ALPinGRB}).}
\label{survProbFigStarburst}
\end{figure}

In Fig.~\ref{survProbFigStarburst} we report both $P_{\rm CP}$ and $P_{\rm ALP}$ (for $B_{\rm ext} = 1 \, \rm nG$ and $B_{\rm ext} < 10^{-15} \, \rm G$) versus ${\cal E}$ for the case of a starburst hosting galaxy and employing the EBL model of~\citet{saldanalopez} by assuming $m_a=10^{-10} \, \rm eV$ and $g_{a\gamma\gamma}=4 \times 10^{-12} \, \rm GeV^{-1}$, which represent the values of the ALP parameter space $(m_a, g_{a\gamma\gamma})$ maximizing $P_{\rm ALP}$ at ${\cal E} = 15 \, \rm TeV$, as shown in Fig.~\ref{parSpaceStarburst}, within the most stringent ALP bounds (\citet{limJulia,mwd}).

Still, even lower $g_{a\gamma\gamma}$ values and different situations -- i.e. both a spiral and a starburst hosting galaxy and both the cases $B_{\rm ext} = 1 \, \rm nG$ and $B_{\rm ext} < 10^{-15} \, \rm G$ -- give rise to a $P_{\rm ALP}$ large enough to allow the observability of GRB 221009A above $10 \, \rm TeV$~(\citet{ALPinGRB,ALPinGRB2}).  

\section{Lorentz invariance violation}

A competitive scenario -- with respect to photon-ALP oscillations -- to explain the TeV detection of GRB 221009A can be represented by the possibility of Lorentz invariance violation (LIV),  predicted by quantum gravity theories, above a threshold energy ${\cal E}_{{\rm LIV}}$ (for a review, see~\citet{addazi}). A consequence of LIV is the modification of the photon dispersion relation, reading
\begin{equation}
{\cal E}^2-p^2 = - \frac{{\cal E}^{n+2}}{{\cal E}_{{\rm LIV}}^n}~,
\label{liv}
\end{equation}
with ${\cal E}$ and $p$ the photon energy and momentum, respectively, and with a resulting modification of the effective photon optical depth (see e.g.,~\cite{tavLIV,gtl}), which is relevant for the TeV detection of GRB 221009A~(\citet{ALPinGRB}) .

\begin{figure}
\resizebox{\hsize}{!}{\includegraphics[clip=true]{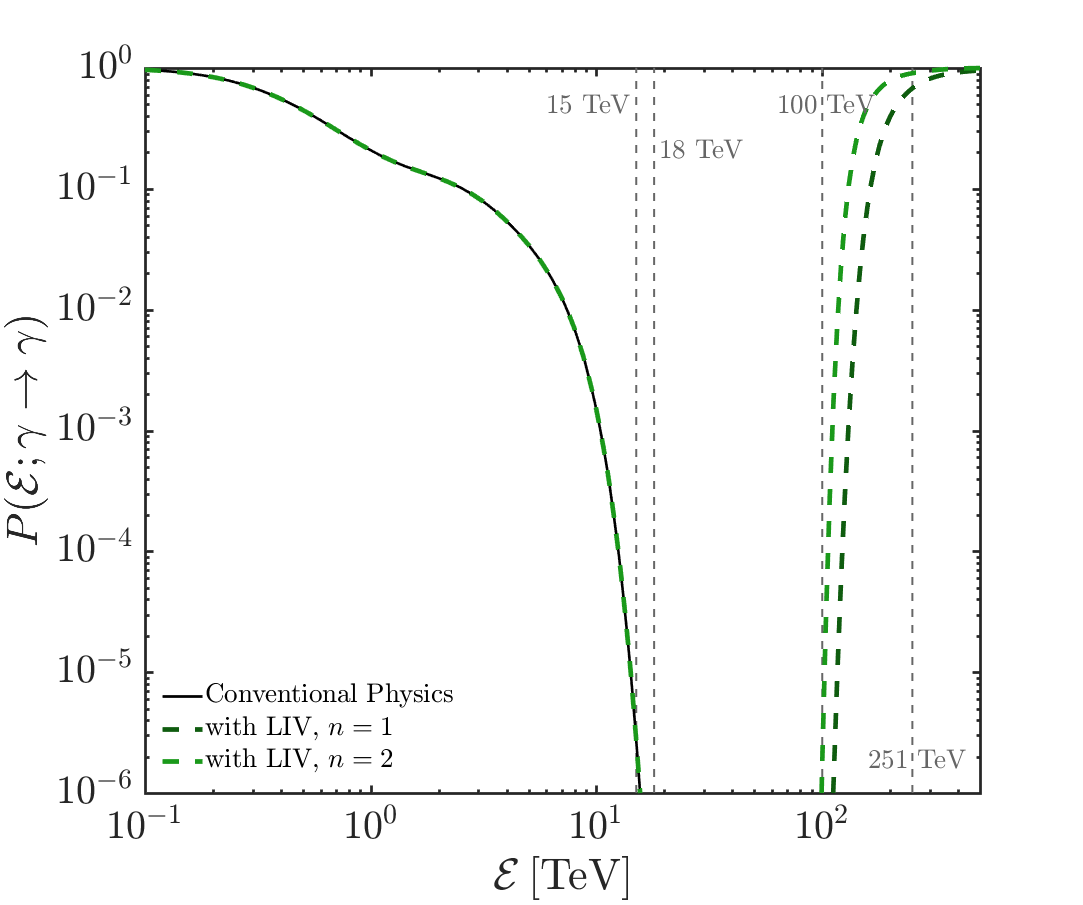}}
\caption{
\footnotesize
Behavior of $P_{\rm CP}$ and of $P_{\rm LIV}$ versus ${\cal E}$. (Credit: adapted from~\citet{ALPinGRB}).}
\label{probLIVfig}
\end{figure}

In Fig.~\ref{probLIVfig} we plot both $P_{\rm CP}$ and the photon survival probability in the presence of LIV effects $P_{\rm LIV}$ by taking ${\cal E}_{{\rm LIV}, n=1} = 3 \times 10^{29} \, {\rm eV}$ in the case $n=1$ and ${\cal E}_{{\rm LIV}, n=2} = 5 \times 10^{21} \, {\rm eV}$ in the case $n=2$, which are values within the current LIV bounds~(\citet{LIVlim}).

From Fig.~\ref{probLIVfig} we infer that LIV is unable to explain the detection by the~\citet{LHAASO}, while it is quite effective to justify that by the~\citet{carpet}, but the latter is less reliable.

\section{Discussion and Conclusions}

Emission models not invoking new physics -- as synchrotron self Compton (SSC) radiation and secondary emission from ultra-high energy protons~(\citet{mirabal,gonzalez,das,zhao,sahu}) -- can hardly explain the detection of ${\cal E} > 10 \, {\rm TeV}$ photons from GRB 221009A  even assuming {\it ad hoc} and often contrived choices of the parameters because of the EBL absorption. So, a tension arises  with the~\citet{LHAASO} detection, when realistic EBL models are taken into account. Moreover, realistic EBL absorption models could become even more severe, as indicated by recent results of the JWST, which suggest a higher infrared background level than currently expected~(\citet{jwst}).

Instead, the photon-ALP oscillation scenario proposed in~\citet{ALPinGRB} (see also~\citet{ALPinGRB2}) turns out to reduce the EBL absorption largely in the energy region of interest, and is extremely effective in justifying both the LHAASO observation and the possible Carpet 2 detection of GRB 221009A within the most stringent ALP bounds~(\citet{limJulia,mwd}). The LIV model proposed in~\citet{ALPinGRB} explains the less reliable Carpet 2 detection of GRB 221009A, but is ineffective at the lower energies observed by the LHAASO collaboration.

In conclusion, although a firm assessment will be possible only when the spectral data of the LHAASO detection are known, we have at least a strong indication of the existence of ALPs coming from GRB 221009A~(\citet{ALPinGRB}) with parameters compatible with two previous ALP hints~(\citet{trgb2012,grdb}) and with values which make ALPs good candidates for cold dark matter~(\citet{arias2012}).
 
\begin{acknowledgements}
We thank Felix Aharonian, Patrizia Caraveo, Elisabete de Gouveia Dal Pino, Giovanni Pareschi, Isabella Prandoni and Franco Vazza for discussions and useful information. The work of G. G. is supported by a contribution from the grant ASI-INAF 2015-023-R.1. The work of M. R. is supported by an INFN grant. This work was made possible also by the funding by the INAF Mini Grant `High-energy astrophysics and axion-like particles', PI: Giorgio Galanti. The work by L. N. is partially supported by the INAF Mini Grant `Shock acceleration in Gamma Ray Bursts', PI: Lara Nava.
\end{acknowledgements}


\begin{thebibliography}{51}

\bibitem[{{Addazi} {et~al.}(2022)}]{addazi} {Addazi}, A., {et~al.} 2022, Progr. in Part. and Nucl. Phys., 125, 103948

\bibitem[{{Agarwal} {et~al.}(2011)}]{ALPpol2} {Agarwal}, N., {Kamal}, A. \& {Jain}, P. 2011, \prd, 83, 065014

\bibitem[{{Arias} {et~al.}(2012)}]{arias2012} {Arias}, P., {et~al.} 2012, JCAP, 06, 013

\bibitem[{{Bassan} {et~al.}(2010)}]{bassan} {Bassan}, N., {Mirizzi}, A. \& {Roncadelli}, M. 2010, JCAP, 05, 010

\bibitem[{{Beck}(2016)}]{SpiralBrev} {Beck}, R. 2016, \aapr, 24, 4

\bibitem[{{Carpet-2 Collaboration}(2022)}]{carpet} {Carpet-2 Collaboration} 2022, ATel \#15669, \url{https://astronomerstelegram.org/?read=15669}

\bibitem[{{CAST Collaboration}(2017)}]{cast} {CAST Collaboration} 2017, Nature Physics, 13, 584

\bibitem[{{Das} \& {Razzaque}(2023)}]{das} {Das}, S. \& {Razzaque}, S. 2023, \aap, 670, 12

\bibitem[{{Day} \& {Krippendorf}(2018)}]{day} {Day}, F. \& {Krippendorf}, S. 2018, Galaxies, 6, 45

\bibitem[{{De Angelis} {et~al.}(2011)}]{dgr2011} {De Angelis}, A., {Galanti}, G. \& {Roncadelli}, M. 2011, \prd, 84, 105030

\bibitem[{{De Angelis} {et~al.}(2007)}]{drm2007} {De Angelis}, A., {Roncadelli}, M. \& {Mansutti}, O. 2007, \prd, 76, 121301

\bibitem[{{Dessert} {et~al.}(2022)}]{mwd} {Dessert}, C., {Dunsky}, D. \& {Safdi}, B. R. 2022, \prd, 105, 103034

\bibitem[{{Dobrynina} {et~al.}(2015)}]{raffelt2015} {Dobrynina}, A., {Kartavtsev}, A. \& {Raffelt}, G. 2015, \prd, 91, 083003

\bibitem[{{Dom\'inguez} {et~al.}(2011)}]{dominguez} {Dom\'inguez}, A., {et~al.} 2011, \mnras, 410, 2556

\bibitem[{{Durrer} \& {Neronov}(2013)}]{durrerneronov} {Durrer}, R. \& {Neronov}, A. 2013 \aapr, 21, 62

\bibitem[{{Dwek} \& {Krennrich}(2013)}]{dwek} {Dwek}, E. \& {Krennrich}, F. 2013, Astropart. Phys., 43, 112

\bibitem[{{Franceschini} \& {Rodighiero}(2017)}]{franceschinirodighiero} {Franceschini}, A. \& {Rodighiero}, G. 2017, \aap, 603, 34 

\bibitem[{{Furlanetto} \& {Loeb}(2001)}]{furlanettoloeb2001} {Furlanetto}, S. \& {Loeb}, A. 2001, \apj, 556, 619

\bibitem[{{Galanti}(2022)}]{galantiTheo} {Galanti}, G. 2022, \prd,105, 083022

\bibitem[{{Galanti}(2023)}]{galantiPol} {Galanti}, G., 2023, \prd, 107, 043006

\bibitem[{{Galanti} {et~al.}(2022a)}]{ALPinGRB} {Galanti}, G., {Nava}, L., {Roncadelli}, M., {Tavecchio}, F. \& {Bonnoli}, G., 2022a, arXiv:2210.05659

\bibitem[{{Galanti} \& {Roncadelli}(2018a)}]{galantironcadelli20118prd} {Galanti}, G. \& {Roncadelli}, M. 2018a, \prd, 98, 043018

\bibitem[{{Galanti} \& {Roncadelli}(2018b)}]{grjhea} {Galanti}, G. \& {Roncadelli}, M. 2018b, J. High Energy Astrophys., 20, 1

\bibitem[{{Galanti} \& {Roncadelli}(2022)}]{universe} {Galanti}, G. \& {Roncadelli}, M. 2022, Universe, 8, 253

\bibitem[{{Galanti} {et~al.}(2020b)}]{grdb} {Galanti}, G., {Roncadelli}, M., {De Angelis}, A. \& {Bignami}, G. F. 2020b, \mnras, 493, 1553    

\bibitem[{{Galanti} {et~al.}(2022b)}]{ALPinGRB2} {Galanti}, G., {Roncadelli}, M. \& {Tavecchio}, F. 2022b, arXiv:2211.06935.

\bibitem[{{Galanti} {et~al.}(2023b)}]{grtBlazar} {Galanti}, G., {Roncadelli}, M. \& {Tavecchio}, F. 2023b, arXiv:2301.08204

\bibitem[{{Galanti} {et~al.}(2023a)}]{grtcClu} {Galanti}, G., {Roncadelli}, M., {Tavecchio}, F. \& {Costa}, E. 2023a, \prd, 107, 103007 

\bibitem[{{Galanti} {et~al.}(2020a)}]{gtl} {Galanti}, G., {Tavecchio}, F. \& {Landoni}, M. 2020a, \mnras, 491, 5268

\bibitem[{{Galanti} {et~al.}(2019)}]{gtre2019} {Galanti}, G., {Tavecchio}, F., {Roncadelli}, M. \& {Evoli}, C. 2019, \mnras, 487, 123

\bibitem[{{Gilmore} {et~al.}(2012)}]{gilmore} {Gilmore}, R. C., {Somerville}, R. S., {Primack}, J. R. \& {Dom\'inguez}, A. 2012, \mnras, 422, 3189

\bibitem[{{Gonzalez} {et~al.}(2023)}]{gonzalez} {Gonzalez}, M. M., {et~al.} 2023, \apj, 944, 178

\bibitem[{{Heisenberg} \& {Euler}(1936)}]{hew1} {Heisenberg}, W. \& {Euler}, H. 1936, Z. Phys., 98, 714

\bibitem[{{Hoyle}(1969)}]{hoyle1969} {Hoyle}, F. 1969, \nat, 223, 936

\bibitem[{{Jaeckel} \& {Ringwald}(2010)}]{JR2010} {Jaeckel}, J. \& {Ringwald}, A. 2010, Ann. Rev. Nucl. Part. Sci., 60, 405

\bibitem[{{Jain} {et~al.}(2002)}]{ALPpol1} {Jain}, P., {Panda}, S. \& {Sarala}, S. 2002, \prd, 66, 085007

\bibitem[{{Jansson} \& {Farrar}(2012a)}]{jansonfarrar1} {Jansson}, R. \& {Farrar}, G. R. 2012a, \apj, 757, 14 

\bibitem[{{Jansson} \& {Farrar}(2012b)}]{jansonfarrar2} {Jansson}, R. \& {Farrar}, G. R. 2012b, \apjl, 761, L11

\bibitem[{{Kohri} \& {Kodama}(2017)}]{kohri2017} {Kohri}, K. \& {Kodama}, H. 2017, \prd, 96, 051701

\bibitem[{{Kronberg} {et~al.}(1999)}]{kronbergleschhopp1999} {Kronberg}, P. P., {Lesch}, H. \& {Hopp}, U. 1999, \apj, 511, 56

\bibitem[{{Labb\'e} {et~al.}(2022)}]{jwst} {Labb\'e}, I., {et~al.} 2022, arXiv:2207.12446

\bibitem[{{Lang} {et~al.}(2019)}]{LIVlim} {Lang}, R. G., {Mart\'inez-Huerta}, H. \& {de Souza}, V. 2019, \prd, 99, 043015

\bibitem[{{Levan} {et~al.}(2023)}]{GRB221009Ahost} {Levan}, A. J., {et~al.} 2023, arXiv:2302.07761

\bibitem[{{LHAASO Collaboration}(2022)}]{LHAASO} {LHAASO Collaboration} 2022, GCN Circular n. 32677, \url{https://gcn.gsfc.nasa.gov/gcn3/32677.gcn3}

\bibitem[{{Lopez-Rodriguez} {et~al.}(2021)}]{LopezRodriguez2021} {Lopez-Rodriguez}, E., {Guerra}, J. A., {Asgari-Targhi}, M. \& {Schmelz}, J. T. 2021, \apj, 914, 24

\bibitem[{{Maiani} {et~al.}(1986)}]{mpz} {Maiani}, L., {Petronzio}, R. \& {Zavattini}, E., 1986, Phys. Lett. B, 175, 359

\bibitem[{{Mirabal}(2023)}]{mirabal} {Mirabal}, N. 2023, \mnras, 519, 85

\bibitem[{{Neronov} \& {Vovk}(2010)}]{neronovvovk} {Neronov}, A. \& {Vovk}, I. 2010, Science, 328, 73

\bibitem[{{Payez} {et~al.}(2011)}]{ALPpol3} {Payez}, A., {Cudell}, J. R. \& {Hutsem\'ekers}, D. 2011, \prd, 84, 085029

\bibitem[{{Perna} {et~al.}(2012)}]{ALPpol5} {Perna}, R., {et~al.} 2012, \apj, 748, 116

\bibitem[{{Pshirkov} {et~al.}(2016)}]{upbbext} {Pshirkov}, M. S., {Tinyakov}, P. G. \& {Urban}, F. R. 2016, \prl, 116, 191302

\bibitem[{{Raffelt} \& {Stodolsky}(1988)}]{raffeltstodolsky} {Raffelt}, G. \& {Stodolsky}, L. 1988, \prd, 37, 1237

\bibitem[{{Rees} \& {Setti}(1968)}]{reessetti1968} {Rees}, M. J. \& {Setti}, G. 1968, \nat, 219, 127

\bibitem[{{Ringwald}(2012)}]{r2012} {Ringwald}, A. 2012, Phys. Dark Univ., 1, 116

\bibitem[{{Sahu} {et~al.}(2023)}]{sahu} {Sahu}, S., {Medina-Carrillo}, B., {S\'anchez-Col\'on}, G. \& {Rajpoot}, S. 2023, \apj, 942, 30

\bibitem[{{Saldana-Lopez} {et~al.}(2021)}]{saldanalopez}  {Saldana-Lopez}, A., {et~al.} 2021, \mnras, 507, 5144 

\bibitem[{{S\'anchez-Conde} {et~al.}(2009)}]{sanchezconde2009} {S\'anchez-Conde}, M. A., {et~al.} 2009, \prd, 79, 123511

\bibitem[{{Schwinger}(1951)}]{hew3} {Schwinger}, J. 1951, Phys. Rev., 82, 664

\bibitem[{{Simet} {et~al.}(2008)}]{simet2008} {Simet}, M., {Hooper}, D. \& {Serpico}, P. D. 2008, \prd, 77, 063001

\bibitem[{{Sisk-Reyn{\'e}s} {et~al.}(2022)}]{limJulia} {Sisk-Reyn{\'e}s}, J., {et~al.} 2022, \mnras, 510, 1264

\bibitem[{{Stargate Collaboration}(2022)}]{redshift} {Stargate Collaboration} 2022, GCN Circular n. 32648, \url{https://gcn.gsfc.nasa.gov/gcn3/32648.gcn3}

\bibitem[{{Tavecchio} \& {Bonnoli}(2016)}]{tavLIV} {Tavecchio}, F. \& {Bonnoli}, G. 2016, \aap, 585, A25

\bibitem[{{Tavecchio} {et~al.}(2015)}]{trg2015} {Tavecchio}, F., {Roncadelli}, M. \& {Galanti}, G. 2015, Phys. Lett. B, 744, 375

\bibitem[{{Tavecchio} {et~al.}(2012)}]{trgb2012} {Tavecchio}, F., {Roncadelli}, M., {Galanti}, G. \& {Bonnoli}, G. 2012, \prd, 86, 085036

\bibitem[{{Thompson} {et~al.}(2006)}]{Thompson2006} {Thompson}, T. A., {et~al.} 2006, \apj, 645, 186

\bibitem[{{Weisskopf}(1936)}]{hew2} {Weisskopf}, V. S. 1936, K. Dan. Vidensk. Selsk. Mat. Fys. Medd., 14, 6

\bibitem[{{Wouters} \& {Brun}(2012)}]{wb2012} {Wouters}, D. \& {Brun}, P. 2012, \prd, 86, 043005

\bibitem[{{Yao} {et~al.}(2017)}]{yaomanchesterwang 2017} {Yao}, J. M., {Manchester}, R. N. \& {Wang}, N. 2017, \apj, 835, 29

\bibitem[{{Zhao} {et~al.}(2023)}]{zhao} {Zhao}, Z.-C., {Zhou}, Y. \& {Wang}, S. 2023, EPJC, 83, 92


\end{thebibliography}
\end{document}